\documentclass[prd,a4paper,twocolumn]{revtex4}
\usepackage{graphicx}
\usepackage[]{subfigure}

\newcommand{\nc}{\newcommand}

\nc{\be}[1]{\begin{equation}\mbox{$\label{#1}$}}
\nc{\bea}[1]{\begin{eqnarray} \mbox{$\label{#1}$}}
\nc{\Section}[2]{\section{#2}\label{#1}}
\nc{\Bibitem}[1]{\bibitem{#1}}
\nc{\Label}[1]{\label{#1}}

\nc{\eea}{\end{eqnarray}}
\nc{\ee}{\end{equation}}

\nc{\bdm}{\begin{displaymath}}
\nc{\edm}{\end{displaymath}}
\nc{\dpsty}{\displaystyle}
\nc{\bc}{\begin{center}}
\nc{\ec}{\end{center}}
\nc{\ba}{\begin{array}}
\nc{\ea}{\end{array}}
\nc{\bab}{\begin{abstract}}
\nc{\eab}{\end{abstract}}
\nc{\btab}{\begin{tabular}}
\nc{\etab}{\end{tabular}}
\nc{\bit}{\begin{itemize}}
\nc{\eit}{\end{itemize}}
\nc{\ben}{\begin{enumerate}}
\nc{\een}{\end{enumerate}}
\nc{\bfig}{\begin{figure}}
\nc{\efig}{\end{figure}}

\nc{\arreq}{&\!=\!&}
\nc{\arrmi}{&\!-\!&}
\nc{\arrpl}{&\!+\!&}
\nc{\arrap}{&\!\!\!\approx\!\!\!&}
\nc{\non}{\nonumber}
\nc{\align}{\!\!\!\!\!\!\!\!&&}

\def\lsim{\; \raise0.3ex\hbox{$<$\kern-0.75em
      \raise-1.1ex\hbox{$\sim$}}\; }
\def\gsim{\; \raise0.3ex\hbox{$>$\kern-0.75em
      \raise-1.1ex\hbox{$\sim$}}\; }

\nc{\DOT}{\hspace{-0.08in}{\bf .}\hspace{0.1in}}
\nc{\Laada}{\hbox {$\sqcap$ \kern -1em $\sqcup$}}
\nc\loota{{\scriptstyle\sqcap\kern-0.55em\hbox{$\scriptstyle\sqcup$}}}
\nc\Loota{{\sqcap\kern-0.65em\hbox{$\sqcup$}}}
\nc\laada{\Loota}
\nc{\qed}{\hskip 3em \hbox{\BOX} \vskip 2ex}

\nc{\real}{{\rm I \! R}}
\nc{\Z}{{\sf Z \!\!\! Z}}
\nc{\complex}{{\rm C\!\!\! {\sf I}\,\,}}
\def\bigid{\leavevmode\hbox{\small1\kern-3.8pt\normalsize1}}
\def\id{\leavevmode\hbox{\small1\kern-3.3pt\normalsize1}}
\nc{\slask}{\!\!\!/}
\nc{\bis}{{\prime\prime}}
\nc{\pa}{\partial}
\nc{\na}{\nabla}
\nc{\ra}{\rangle}
\nc{\la}{\langle}
\nc{\goto}{\rightarrow}
\nc{\swap}{\leftrightarrow}

\nc{\EE}[1]{ \mbox{$\cdot10^{#1}$} }
\nc{\abs}[1]{\left|#1\right|}
\nc{\at}[2]{\left.#1\right|_{#2}}
\nc{\norm}[1]{\|#1\|}
\nc{\abscut}[2]{\Abs{#1}_{\scriptscriptstyle#2}}
\nc{\vek}[1]{{\rm\bf #1}}
\nc{\integral}[2]{\int\limits_{#1}^{#2}}
\nc{\inv}[1]{\frac{1}{#1}}
\nc{\dd}[2]{{{\partial #1}\over{\partial #2}}}
\nc{\ddd}[2]{{{{\partial}^2 #1}\over{\partial {#2}^2}}}
\nc{\dddd}[3]{{{{\partial}^2 #1}\over
    {\partial #2 \partial #3}}}
\nc{\dder}[2]{{{d #1}\over{d #2}}}
\nc{\ddder}[2]{{{d^2 #1}\over{d {#2}^2}}}
\nc{\dddder}[3]{{d^2 #1}\over
    {d #2 d #3}}
\nc{\dx}[1]{d\,^{#1}x}
\nc{\dy}[1]{d\,^{#1}y}
\nc{\dz}[1]{d\,^{#1}z}
\nc{\dl}[1]{\frac{d\,^{#1}l}{(2\pi)^{#1}}}
\nc{\dk}[1]{\frac{d\,^{#1}k}{(2\pi)^{#1}}}
\nc{\dq}[1]{\frac{d\,^{#1}q}{(2\pi)^{#1}}}

\nc{\bfT}{{\bf T }}

\nc{\cA}{{\cal A}}
\nc{\cB}{{\cal B}}
\nc{\cD}{{\cal D}}
\nc{\cE}{{\cal E}}
\nc{\cG}{{\cal G}}
\nc{\cH}{{\cal H}}
\nc{\cL}{{\cal L}}
\nc{\cO}{{\cal O}}
\nc{\cT}{{\cal T}}
\nc{\cN}{{\cal N}}
\nc{\cR}{{\cal R}}
\nc{\rvac}[1]{|{\cal O}#1\rangle}
\nc{\lvac}[1]{\langle{\cal O}#1|}
\nc{\rvacb}[1]{|{\cal O}_\beta #1\rangle}
\nc{\lvacb}[1]{\langle{\cal O}_\beta #1 |}
\nc{\bb}{\bar{\beta}}
\nc{\bt}{\tilde{\beta}}
\nc{\ctH}{\tilde{\cal H}}
\nc{\chH}{\hat{\cal H}}
\nc{\1}{\aa}
\nc{\2}{\"{a}}
\nc{\3}{\"{o}}
\nc{\4}{\AA}
\nc{\5}{\"{A}}
\nc{\6}{\"{O}}
\nc{\al}{\alpha}
\nc{\g}{\gamma}
\nc{\Del}{\Delta}
\nc{\e}{\textrm{e}}
\nc{\eps}{\epsilon}
\nc{\lam}{\lambda}
\nc{\Om}{\Omega}
\nc{\ve}{\varepsilon}
\nc{\mn}{{\mu\nu}}
\nc{\vp}{\varphi}

\nc{\rf}[1]{(\ref{#1})}
\nc{\nn}{\nonumber \\*}
\nc{\bfB}{\bf{B}}
\nc{\bfv}{\bf{v}}
\nc{\bfx}{\bf{x}}
\nc{\bfy}{\bf{y}}
\nc{\vx}{\vec{x}}
\nc{\vy}{\vec{y}}
\nc{\oB}{\overline{B}}
\nc{\oI}{\overline{I}}
\nc{\oR}{\overline{R}}
\nc{\rar}{\rightarrow}
\nc{\ti}{\times}
\nc{\slsh}{\hskip-5pt/}
\nc{\sm}{Standard~Model~}
\nc{\MP}{M_{\rm Pl}}
\nc{\mpl}{M_{\rm Pl}}
\nc{\tp}{t_{\rm Pl}}

\nc{\pmin}{p_{\rm min}}
\nc{\pmax}{p_{\rm max}}
\nc{\fo}{f_0}
\nc{\foi}{f_{0,i}\,}
\nc{\fop}{f_0^P}
\nc{\fou}{f_0^U}

\nc{\eff}{{\rm eff}}
\nc{\MT}{M_{\rm T}}
\nc{\ML}{M_{\rm L}}
\nc{\kk}{\vek{k}}
\nc{\pp}{{\rm p}}
\nc{\pt}{\partial_t}
\nc{\half}{{1\over 2}}
\nc{\w}{\omega}
\nc{\uhat}{\hat{U}_\w}

\nc{\etal}{\mbox{\it et al.}}
\nc{\ie}{{\it i.e. }}
\nc{\eg}{{\it e.g. }}
\nc{\trh}{T_{\rm RH}}
\nc{\ad}{{a'\over a}}
\nc{\bd}{{b'\over b}}
\nc{\Rd}{{R'\over R}}
\nc{\diag}{{\textrm{diag}}}
\nc{\mato}[1]{\tilde{#1}}
\nc{\sinn}{\textrm{sinn}}
\nc{\sech}{\textrm{sech}}
\nc{\I}{\textrm{I}}
\nc{\II}{\textrm{II}}
\nc{\III}{\textrm{III}}
\nc{\vev}[1]{\langle #1 \rangle}
\nc{\hyp}{\,\; F_{1{\hskip -16pt}2}{\hskip 11pt}}
\nc{\brhom}{\overline{\rho}_M}
\nc{\brho}{\overline{\rho}}
\nc{\rhob}{\overline{\rho}}
\nc{\Pb}{\overline{P}}
\nc{\bH}{\overline{H}}
\nc{\ep}{{1+4\eps}}

\def\smiley{\hbox{\large$\bigcirc$\hspace{-.80em}%
\raise.2ex\hbox{$\cdot\cdot$}\kern-.61em    
\lower.2ex\hbox{\scriptsize$\smile$}}\ }

\def\frowney{\hbox{\large$\bigcirc$\hspace{-.80em}%
\raise.2ex\hbox{$\cdot\cdot$}\kern-.635em
\lower.2ex\hbox{\scriptsize$\frown$}}\ }

\begin{document}

\title{Bayesian analysis of Friedmannless cosmologies}

\author{{\O}. Elgar\o y}
\email{oelgaroy@astro.uio.no}
\affiliation{Institute of theoretical astrophysics, University of Oslo,Box 1029, 0315 Oslo, NORWAY}
\author{T. Multam\"{a}ki}
\email{tuomul@utu.fi}
\affiliation{Department of Physics, University of Turku, FIN-20014 Turku, FINLAND}
\date{\today}

\begin{abstract}
Assuming only a homogeneous and isotropic universe and using both the 
`Gold' Supernova Type Ia sample of Riess et al. and the results from 
the Supernova Legacy Survey, we calculate the 
Bayesian evidence of a range of different parameterizations of 
the deceleration parameter.  We consider both spatially flat and 
curved models.  Our results show that although there is strong 
evidence in the data for an accelerating universe, there is little 
evidence that the deceleration parameter varies with redshift.  
\end{abstract}
\maketitle

\section{Introduction}   

The accelerating expansion of the universe is probably the most important 
discovery in cosmology in the last decade.  Direct evidence for 
cosmic acceleration comes from the Hubble diagram with supernovae 
of type Ia (SNIa) as standard (or standardizable) candles 
\cite{perlmutter,riess,snls}, and indirect support comes from 
e.g. observations of the large-scale distribution of matter 
\cite{tegmark,cole} combined with measurements of temperature 
anisotropies in the cosmic microwave background (CMB) radiation 
\cite{spergel}. 

While the most straightforward 
interpretation of the acceleration is that the energy density of the 
universe is presently dominated by the cosmological constant, the 
problem of understanding why it is so small compared with the natural 
energy scale of quantum gravity has led to proposals of a jungle of 
alternatives.  The proposals can be roughly divided into two 
classes: modifications of the right-hand side or the left-hand side 
of the Einstein equations.  The first case corresponds to changing the 
theory of gravity from standard general relativity (GR), while in 
the second case some kind of negative-pressure fluid is added to the 
energy-momentum tensor.  At present none of these possibilities 
can be ruled out.   
The literature is vast, but useful reviews 
are found in  references \cite{peebles,paddy1}.

In this paper we ask the question of what one can learn from the 
SNIa data about the kinematics of the universe.  This sidesteps 
the issue of the cause of the acceleration and leads to fairly 
model-independent conclusions since we only assume that the universe 
is homogeneous and isotropic and is described by a metric.  We think 
a study like this is timely since sometimes one sees in the literature 
wide-ranging conclusions drawn on the basis of fitting specific 
models to SNIa data.  For example, the recent discussion of 
`phantom energy' \cite{caldwell} is based on the fact that fits 
with a dark energy component with a constant equation of state parameter 
$w$ prefer values less than $-1$.  We think it is important to bear in 
mind what exactly the SNIa data is telling us about the expansion of the 
universe when assessing the merit of claims like this. 

Our study is similar in spirit to that of a recent paper by Shapiro and Turner
\cite{shapiro}, 
but our analysis is based on the concept of the Bayesian evidence.  
Also, we analyze both the `Gold' SNIa sample of Riess et al. and the more recent sample from 
the Supernova Legacy Survey (SNLS), and show that they lead to slightly 
different conclusions about the expansion history of the universe, 
although none of the data sets can provide strong constraints.

\subsection{Friedmannless Cosmology}
Assuming homogeneity and isotropy on large scales, the requirement 
for a maximally symmetric subspace leads us to the 
Friedmann-Robertson-Walker (FRW) metric:
\be{frw}
ds^2=dt^2-a^2(t)\Big[\frac{dr^2}{1-k r^2}+r^2d\Omega^2\Big],
\ee
where $a(t)$ is the scale factor. Note that we have not assumed
spatial flatness, $k=0$, in order to keep the discussion as 
general as possible. 

The deceleration parameter is defined as
\be{dec}
q\equiv -\frac{1}{H^2}\frac{\ddot{a}}{a}=
\frac 12 (1+z)\frac{(H(z)^2)'}{H(z)^2}-1,
\ee
where we have written $q$ in terms of the Hubble parameter and 
$'\equiv d/dz$. 
Similarly, the jerk is given by:
\be{jerk}
j\equiv -\frac{1}{H^3}\frac{\dddot{a}}{a}=-\Big[\frac 12 (1+z)^2 
\frac{(H^2)''}{H^2}-(1+z)\frac{(H^2)'}{H^2}+1\Big].
\ee
Naturally, one may consider expansions of $q(z)$, $j(z)$ or even
higher order parameters. However, the order of the parameter, or
the number of derivatives, is related to the number
of free parameters in $H(z)$. This is clearly demonstrated by
considering for example models with constant deceleration
parameter, $q=q_0$, or jerk, $j=j_0$. In this case equations
(\ref{dec}) and (\ref{jerk}) are straightforwardly solvable
with solutions of the form
\bea{sols}
H^2(z) & = & c_1(1+z)^{2(1+q_0)}\\
H^2(z) & = & c_1(1+z)^{\alpha_1}+c_2(1+z)^{\alpha_2},
\eea
where 
\be{jerksol2}
\alpha_{1,2}\equiv\frac 32 \pm\sqrt{2(-j_0-1)+\frac 94}.
\ee
Hence, expanding the jerk can be viewed as requiring more
parameters, or cosmic fluids, than expanding the 
deceleration parameter. Guided by these considerations
we consider expansions of $q(z)$. This furthermore allows us
to compare our results with those of \cite{shapiro} who 
reach their conclusions by a different method.

In a standard matter only universe, Einstein-de Sitter (EdS) model,
the deceleration parameter is simply $q=1/2$ and jerk $j=-1$.
A $\Lambda$CDM universe with $H^2=H_0^2(\Omega_m(1+z)^3+
\Omega_\Lambda)$ on the other hand has a non-constant 
deceleration parameter,
\be{declcdm}
q(z)=-\frac{\Omega_\Lambda/\Omega_m-\frac 12 (1+z)^3}{\Omega_\Lambda/\Omega_m+(1+z)^3},
\ee
and again a constant jerk $j=-1$. This also demonstrates
how mapping from cosmological models to cosmological parameters
is not unique, eg. both EdS and $\Lambda$CDM models have the same
jerk further supporting the use of $q$ as an expansion parameter.

We make minimal or no assumptions about the matter content of
the universe. Considering the deceleration parameter we can avoid
the question of the gravitational theory, whether standard Einstein's
equations hold or not, ie. we make no assumptions on the connection
between the energy density of matter and the evolution of the universe.
Similar constructions have been considered previously: Dvali and Turner 
\cite{dtmodel} considered adding a correction to the Friedmann
equation, Freese \& Lewis \cite{Freese} considered generalized Friedmann
equations in their Cardassian models and generalizations
in a similar spirit were also considered in a previous paper \cite{osterix}.

In addition to considering straightforward linear expansions of the deceleration parameter, $q(z)=\sum q_iz^i$, we also study more physically 
guided expansions, $q(z)=\sum q_i (1+z)^{-3i}$. 
Such expansions are motivated by the fact that if we assume that the 
energy density is conserved in the expanding universe, 
cold dark matter density is also conserved and hence $\rho_m\sim a^{-3}\sim
 (1+z)^3$. One can view this requirement simply as arising from 
conservation of the
number of the particles and not as a property of the Einstein's
equations. Energy conservation for example holds in a large class of 
modified gravity models, so called  $f(R)$ models 
both within the metric and Palatini approach \cite{tomi}. 
As example, consider the $\Lambda$CDM model, which when 
expanded in powers of $(1+z)^{-3}$ gives
\be{lcdmexp}
q\approx \frac 12 -\frac 32\frac{\Omega_\Lambda}{\Omega_m}(1+z)^{-3}
+\frac{3}{2}\left(\frac{\Omega_\Lambda}{\Omega_m}\right)^2(1+z)^{-6}.
\ee

The models we consider along with the very conservative priors
are shown in table \ref{tab:models}.

\begin{table}[ht]
\begin{center}
\begin{tabular}{l|l|l}
\hline Model &  & Priors\\
\hline 
$M_0$ & $q=q_0$	& $q_0\in[-5,5]$\\
$M_1$ & $q=q_0+q_1z$	&
$\left\{\begin{array}{l} q_0\in[-5,5]\\ q_1\in[-5,5] \end{array} \right.$\\
$M_2$ & $q=q_0+q_1z+q_2z^2$&
$\left\{\begin{array}{l} q_0\in[-5,5]\\ q_1\in[-5,5]\\ q_2\in[-5,5] \end{array} \right.$\\
$M_3$ & $q=q_0+q_1/(1+z)^3$ & 
$\left\{\begin{array}{l} q_0\in[-5,5]\\ q_1\in[-5,5] \end{array} \right.$\\
$M_4$ & $q=q_0+q_1/(1+z)^3+q_2/(1+z)^6$ & 
$\left\{\begin{array}{l} q_0\in[-5,5]\\ q_1\in[-5,5]\\ q_2\in[-5,5] \end{array} \right.$\\
$M_5$ & 
$q=
\left\{
\begin{array}{ll}
q_0, & z\leq z_t\\
q_1, & z>z_t\\
\end{array}
\right.
$ & $\left\{\begin{array}{l} q_0\in[-5,5]\\ q_1\in[-5,5]\\ z_t\in[0,1] \end{array} \right.$\\
\hline Non-flat &  & \\
\hline 
$M_6$ & $q=q_0$	& 
$\left\{\begin{array}{l} q_0\in[-5,5]\\ k=\{-1,1\} \end{array} \right.$\\
$M_7$ & $q=q_0+q_1z$ &
$\left\{\begin{array}{l} q_0\in[-5,5]\\ q_1\in[-5,5]\\ k=\{-1,1\} \end{array} \right.$
\end{tabular}
\end{center}
\caption{Model parameters and priors}
\label{tab:models}
\end{table}

\subsection{Bayesian Evidence}
The Bayesian evidence (BE), or $E(M)$, 
is defined as the probability of the data $D$
given the model $M$ with a set of parameters $\theta$,
\be{be}
E(M)\equiv P(D|M)=\int d\theta\, P(D|\theta,M)P(\theta|M),
\ee
where $P(\theta|M)$ is the prior on the set of parameters, normalized
to unity. We follow here the common assumption that the parameter priors
are top hat and hence (\ref{be}) can be written as
\be{be2}
E(M)=\frac{\int d\theta\, P(D|\theta,M)}
{\int d\theta}.
\ee
In this paper we approximate the probability by
$P(D|\theta)\approx \exp(-\chi^2(\theta)/2)$. Since we are
interested in comparing models, we leave the normalization arbitrary.

In addition to BE, we will at times refer to an unnormalized
Bayesian evidence, which is simply
\be{ube}
\widetilde{E}(M)=\int d\theta\, P(D|\theta,M).
\ee

In order to compare models, we utilize the Bayes Factor
which is defined as the ratio of evidences of two models $M_i$ and $M_j$:
\be{bef}
B_{ij}=\frac{E(M_i)}{E(M_j)}.
\ee
If $B_{ij}>1$, model $M_i$ is preferred over $M_j$, given the data.
In similar spirit as 
in the frequentist approach, where one compares the $\chi^2$s of
given models and concludes how significantly a model fits the data better,
in the Bayesian Evidence framework one can assign significance to the
difference between models by using the Jeffreys Scale \cite{jeffreys} 
as shown in Table \ref{js}.
\begin{table}[ht]
\begin{center}
\begin{tabular}{l|l}
\hline
Log of Bayes Factor & Significance\\
\hline
$\ln B_{ij}<1$ & Not significant\\
$1< \ln B_{ij}<2.5 $ & Substantial\\
$2.5< \ln B_{ij}<5 $ & Strong\\
$5< \ln B_{ij}$ & Decisive
\end{tabular}
\end{center}
\caption{Jeffreys Scale}\label{js}
\end{table}
Note that $B_{ij}$ is assumed to be greater than one ie. model $M_i$
has larger evidence than model $M_j$.

An approximation to the Bayesian Evidence is the Bayesian Information
Criterion (BIC) \cite{schwartz, bicandrew}. BIC also penalizes for
extra parameters by reducing the likelihood,
\be{bic}
BIC=-2\cL+k\, \ln N,
\ee
where $k$ is the number of parameters and $N$ is the number of data points
used in the fit. Approximating the likelihood by $\chi^2$, it is
evident how in the frequentist language the fit to a model
with more parameters needs to be substantially better in order
to justify the extended model. For example, for a SNIa dataset with 
157 supernovae, a two parameter model must have a significantly
better fit to the data, $\Delta\chi^2\approx 10$, than a single
parameter model in order to justify the more extended model 
($\Delta BIC\approx 5$).


\section{Methods}

In this work we use data from two SNIa compilations, the 
Riess Gold set \cite{riess} with 157 supernovae up to $z=1.76$
and the more recent SNIa data released by the Supernova Legacy Survey 
(SNLS) \cite{snls} containing 115 supernovae with $z<1$.

Within the Friedmannless Cosmology approach, the geometry of the universe
becomes relevant at this stage since the luminosity distance
depends on the spatial curvature $k$. The dimensionless luminosity distance 
is
\be{dl}
d_L(z)=(1+z)\sinn_k\Big(\int_0^z\frac{du}{H(u)}\Big),
\ee
where $\sinn_k(x)$ is the commonly used function
\be{sinn}
\sinn_{k}(x)=\left\{\begin{array}{l}
\sin(x),\ k=+1\\
x,\ k=0\\
\sinh(x),\ k=-1
\end{array}
\right.
\ee

The two surveys use somewhat different methods for calculating the
$\chi^2$ of a model (see eg. \cite{riess, snls, nesseris} for a more 
 detailed description).
For the Gold set, the $\chi^2$ of a model $M$ with a set of 
parameters $\{\alpha_i\}$ is
\be{goldana}
\chi^2(\{\alpha_i\},\mu_0)=\sum_i^n\frac{(\mu_{obs,i}-5 \log_{10} d_L(z_i;\{\alpha_i\})-\mu_0)^2}{\sigma_i^2},
\ee
where $d_L$ is the Hubble free luminosity distance. The Hubble parameter and 
corrections to the absolute magnitude are hidden in $\mu_0$.

The analysis for the SNLS dataset is somewhat more complicated
due to the appearance of two extra nuisance parameters $\alpha$ and $\beta$:
\begin{eqnarray}
\chi^2(\{\alpha_i\},M+\mu_0,\alpha,\beta)&= & \nonumber \\
\sum_i^n\frac{(\mu_{obs,i}-5 \log_{10} d_L(z_i;\{\alpha_i\})-\mu_0)^2}{\sigma_i^2},
\label{eq:snlsana}
\end{eqnarray}
where now $\sigma_i=\sigma_i(\alpha,\beta)$. The computation of $\chi^2$
using the SNLS is hence computationally more intensive as there are three
nuisance parameters to marginalized over (again, we refer to \cite{snls}
for a detailed description of the procedure).

An often utilized approximation in marginalizing over nuisance parameters
is to marginalize by maximizing the likelihood. 
For example, instead of marginalizing over $\mu_0$ by integrating 
over $\mu_0$ with a given prior, a commonly used approximation
when using the Gold set in studying cosmological models is to
maximize the likelihood by minimizing $\chi^2$ with respect to $\mu_0$.
This is well motivated as can be seen by first
expanding (\ref{goldana}), which gives
\be{goldana2}
\chi^2(\{\alpha_i\},\mu_0)=c_1-2c_2\mu_0+c_3\mu_0^2,
\ee
where $c_i$ are independent of $\mu_0$:
\bea{cs}
c_1 & = & \sum_i^n\frac{(\mu_{obs,i}-5\log_{10}d_L(z_i;\{\alpha_i\}))^2}{\sigma_i^2}\nonumber\\
c_2 & = & \sum_i^n\frac{\mu_{obs,i}-5\log_{10}d_L(z_i;\{\alpha_i\})}{\sigma_i^2}\nonumber\\
c_3 & = & \sum_i^n\sigma_i^{-2}.
\eea
The minimization over $\mu_0$ is trivial, $\mu_0=c_2/c_3$. 
Hence, instead of finding the minimum
of $\chi^2(\{\alpha_i\},\mu_0)$ we can minimize 
$\tilde{\chi}^2(\{\alpha_i\})\equiv \chi^2(\{\alpha_i\},c_2/c_3)$.
In calculating the confidence contours, however, 
this method is only approximate unless we assume no prior on $\mu_0$. 
In the special case with no prior on $\mu_0$, one can carry out the integral explicitly resulting 
in a constant term that will not have an effect in calculating the 
confidence contours. 



As a check of our method, we have confirmed that the 
confidence contours produced using the Gold sample for the two epoch model, 
$M_5$, agree with those presented in \cite{shapiro}.  For the SNLS sample, 
we checked that we reproduced the results in \cite{snls} for flat 
and non-flat $\Lambda{\rm CDM}$ models.


In calculating the Bayesian evidence $E(M)$ of a given model,
integration over the parameters is required. In order to 
assess the validity of the approximation of marginalizing over
nuisance parameters by maximizing the likelihood
we calculate $E(M)$ for the Gold set both by using explicit
integration over $\mu_0$ and by the approximate method.
The parameter priors are chosen to be 
$q_0\in[-1,1],\ q_1\in[-2,2],\ \mu_0\in [42.38,43.89]$
and the comparison is carried out for the constant $q$ and linear $q$ model.
When marginalization over $\mu_0$ is
done explicitly when calculating BE, we expect the resulting evidence
to be less than when $\mu_0$ is marginalized by maximizing the
likelihood at each point. This is due to normalization and to the fact
that the likelihood at a point in parameter space is typically
less when not explicitly maximized. This proves to be the case,
eg. for the constant model 
$E(M_0)$ is about $30$ times larger 
when the marginalization is carried out approximatively.
However, in using BE we are interested in comparing models
and hence a more interesting issue is to compare the two
different models with both marginalization methods. 
We find that the difference between marginalization
schemes is not significant when comparing models,
$B_{01}\approx 2.5$ in both cases. Hence, marginalizing over 
$\mu_0$ by maximizing the likelihood appears to be a reasonable approximation 
in calculating the Bayesian Evidence at least for the parameter 
priors used in this paper.


\begin{table}[ht]
\begin{center}
\begin{tabular}{l|ccl}
\hline 
Model  & $\ln B_{0i}$ & $\chi^2$ & Best fit parameters\\
\hline 
$M_0$ & $0$ & 182.8 & $q_0=-0.29$\\
$M_1$ & $2.0$ & 175.4 & 
$\left\{\begin{array}{l} q_0=-0.70\\ q_1=1.5 \end{array} \right.$\\
$M_2$ & $1.7$ & 173.8 & 
$\left\{\begin{array}{l} q_0=-1.1\\ q_1=4.5\\ q_2=-3.3 \end{array} \right.$\\
$M_3$ & $2.7$ & 174.0 &
$\left\{\begin{array}{l} q_0=0.70\\ q_1=-1.8\end{array} \right.$\\
$M_4$ & $2.6$ & 172.7 &
$\left\{\begin{array}{l} q_0=-0.30\\ q_1=3.2\\ q_2=-4.9 \end{array} \right.$\\
$M_5$ & $1.8$ & 172.4 & 
$\left\{\begin{array}{l} q_0=1.6\\ q_1=0.20\\ z_t=0.11 \end{array} \right.$\\
\hline
Non-flat\\
\hline
$M_6,\ (k=-1)$ & $3.5$ & 176.2 & $q_0=-0.64$\\
$M_6,\ (k=1)$ & $<0$ & 191.1 & $q_0=-0.04$\\
$M_7,\ (k=-1)$ & $1.9$ & 176.2 &
$\left\{\begin{array}{l} q_0=-0.70\\ q_1=0.20\end{array} \right.$\\
$M_7,\ (k=1)$ &  $1.5$ & 176.0 &
$\left\{\begin{array}{l} q_0=-0.70\\ q_1=2.3\end{array} \right.$
\end{tabular}
\end{center}
\caption{Bayes factor relative to the constant $q$ model, minimum $\chi^2$ and best
fit parameters for different models}
\label{tab:results}
\end{table}

\section{Results and discussion}
\subsection{Gold sample}
The Bayes factor relative to the flat constant $q$ model, $\ln B_{0i}$, 
minimum $\chi^2$ and best fit
parameters for the different models are given in table \ref{tab:results}.
Keeping in mind the Jeffrey's scale, we see from the table
that most models are not significantly better than the baseline model $M_0$, 
which is the constant $q$ flat universe model. Only flat models
for which we find strong evidence are those 
 where we
have assumed a more physically motivated expansion
(note that $\Delta \ln B_{ij}=5$ corresponds to odds of $1/13$).
Interestingly, the model with the largest evidence is a constant
$q$ closed universe, which is strongly preferred over its flat universe
counterpart.

In calculating the Bayesian Evidence we use a simple grid. We have 
checked that the results are insensitive to further refinement of the grid.
Note that in contrast the best fit parameters maybe sensitive to grid spacing,
for example due to the existence of nearly degenerate minima. This 
further advocates the use of Bayesian Evidence as an efficient 
model selection tool.



\subsection{The SNLS sample}

Carrying out the same calculations for the SNLS sample of SNIa gives 
the results in table \ref{tab:resultsSNLS}.  The nuisance parameters 
$\alpha$, $\beta$, and $M+\mu_0$ were in each case fairly tightly constrained by the data, so we carried out the marginalization over them 
by fixing them to their best-fit values.   
The model parameters were integrated over with the priors given in table I. 
An unsatisfactory aspect of this approach is that, e.g. the likelihood 
function for the $M_2$ model does not reduce to that of $M_1$ for 
$q_2=0$, or to that of $M_0$ for $q_1=q_2=0$.  We can get a sense of 
how this affects the results by using the best-fit $\alpha$, $\beta$ 
and $M+\mu$ for $M_0$ for {\it all} the models.  The results for this case 
are given in table \ref{tab:resultsSNLS2}. 
Note that the Bayes factors do not change much except for models 
$M6(k=-1)$ and $M6(k=+1)$.  
\begin{table}[ht]
\begin{center}
\begin{tabular}{l|cccl}
\hline 
Model  & $\ln B_{0i}$ & $\chi^2$ & Best fit parameters\\\hline 
$M_0$ & $0$ & 112.0 & $q_0=-0.42$\\
$M_1$ & $1.7$ & 111.2 & 
$\left\{\begin{array}{l} q_0=-0.47\\ q_1=0.24 \end{array} \right.$\\
$M_2$& $0.6$ & 110.5 & 
$\left\{\begin{array}{l} q_0=-0.60\\ q_1=-0.60\\ q_2=0.61 \end{array} \right.$\\
$M_3$ & $2.4$ & 111.8 &
$\left\{\begin{array}{l} q_0=-0.66\\ q_1=0.43\end{array} \right.$\\
$M_4$ & $1.3$ & 110.3 &
$\left\{\begin{array}{l} q_0=-0.23\\ q_1=0.29\\ q_2=-0.88 \end{array} \right.$\\
$M_5$ & $4.2$ & 112.5 & 
$\left\{\begin{array}{l} q_0=-0.60\\ q_1=0.49\\ z_t=0.39 \end{array} \right.$\\
\hline
Non-flat\\
\hline
$M_6,\ (k=-1)$ & $1.8$ & 112.1 & $q_0=-0.21$\\
$M_6,\ (k=+1)$ & $1.1$ & 112.6 & $q_0=-0.68$\\
$M_7,\ (k=-1)$ & $2.0$ & 110.8 &
$\left\{\begin{array}{l} q_0=-0.28\\ q_1=0.27\end{array} \right.$\\
$M_7,\ (k=+1)$ & $1.6$ & 112.4 &
$\left\{\begin{array}{l} q_0=-0.70\\ q_1=2.3\end{array} \right.$
\end{tabular}
\end{center}
\caption{Bayes factor relative to the constant $q$ model, minimum $\chi^2$ and best
fit parameters for different models from the SNLS sample.}
\label{tab:resultsSNLS}
\end{table}
\begin{table}[ht]
\begin{center}
\begin{tabular}{l|cccl}
\hline 
Model  & $\ln B_{0i}$ & $\chi^2$ & Best fit parameters\\\hline 
$M_0$ & $0$ & 112.0 & $q_0=-0.42$\\
$M_1$ & $1.7$ & 111.7 & 
$\left\{\begin{array}{l} q_0=-0.55\\ q_1=0.70 \end{array} \right.$\\
$M_2$& $0.6$ & 111.3 & 
$\left\{\begin{array}{l} q_0=-0.32\\ q_1=-2.01\\ q_2=4.64 \end{array} \right.$\\
$M_3$ & $2.0$ & 111.9 &
$\left\{\begin{array}{l} q_0=-0.11\\ q_1=-0.47\end{array} \right.$\\
$M_4$ & $1.4$ & 111.4 &
$\left\{\begin{array}{l} q_0=-1.06\\ q_1=-4.97\\ q_2=3.73 \end{array} \right.$\\
$M_5$ & $4.1$ & 111.3 & 
$\left\{\begin{array}{l} q_0=-0.46\\ q_1=0.35\\ z_t=0.50 \end{array} \right.$\\
\hline
Non-flat\\
\hline
$M_6,\ (k=-1)$ & $3.3$ & 115.2 & $q_0=-0.24$\\
$M_6,\ (k=+1)$ & $3.4$ & 112.2 & $q_0=-0.65$\\
$M_7,\ (k=-1)$ & $1.7$ & 111.5 &
$\left\{\begin{array}{l} q_0=-0.53\\ q_1=1.53\end{array} \right.$\\
$M_7,\ (k=+1)$ & $1.5$ & 111.6 &
$\left\{\begin{array}{l} q_0=-0.51\\ q_1=-0.73\end{array} \right.$
\end{tabular}
\end{center}
\caption{Bayes factor relative to the constant $q$ model, minimum $\chi^2$ and best
fit parameters for different models from the SNLS sample.  The 
nuisance parameters $\alpha$, $\beta$ and $M+\mu_0$ were marginalized over 
by fixing them to their best-fit values for the $M0$ model.}
\label{tab:resultsSNLS2}
\end{table}

\subsection{Model comparison}
We can now rank the models based on their Bayes factor relative to $M_0$.  
Both the Gold sample and the SNLS provide substantial evidence against 
all models from 
places 3 up to and including 7, and strong evidence against models 
at ranks 8-10, but they do not agree on which models fall into which 
category.  However, the best model in both cases has 
$q(z)=\rm constant$.  It therefore seems fair to conclude 
that there is no significant evidence in the present supernova 
data for a transition from deceleration to acceleration, and claims 
to the contrary are most likely an artifact of the parameterization 
used in the fit to the data.  
Comparing the ranking according to the BIC gives very 
different results for both data sets, and thus one should avoid 
approximations like this in Bayesian model selection problems.   
\begin{table}[ht]
\begin{center}
\begin{tabular}{l|ccccl}
\hline 
Rank  & Gold sample& SNLS & Gold sample (BIC) & SNLS (BIC) \\ \hline 
1     & $M_6(k=+1)$ & $M_0$ & $M_6(k=-1)$ & $M_6(k=+1)$    \\ 
2     & $M_0$       & $M_2$ & $M_3$       & $M_0$          \\ 
3     & $M_7(k=+1)$ & $M_4$ & $M_1$       & $M_6(k=-1)$    \\ 
4     & $M_2$       & $M_7(k=+1)$ & $M_7(k=+1)$ &$M_7(k=+1)$\\ 
5     & $M_5$       & $M_1$       & $M_7(k=-1)$&$M_7(k=-1)$\\ 
6     & $M_7(k=-1)$ & $M_7(k=-1)$ & $M_5$      &$M_1$      \\ 
7     & $M_1$       & $M_3$       & $M_0$,$M_4$&$M_3$      \\ 
8     & $M_4$       & $M_6(k=-1)$ & $M_2$      &$M_2$,$M_5$\\ 
9     & $M_3$       & $M_6(k=+1)$ & $M_6(k=+1)$&$M_4$  \\ 
10    & $M_6(k=-1)$ & $M_5$       &            &  \\ 
\hline  
\end{tabular}
\end{center}
\caption{The models ranked according to $B_{0i}$ for the Gold sample and 
the SNLS sample, and ranked according to the Bayesian Information Criterion.}
\label{tab:ranking}
\end{table}

That the two SNIa samples prefer slightly different models has 
already been noted in, e.g., \cite{jassal} where it was shown that 
the SNLS best-fit model was in concordance with the best-fit 
WMAP model \cite{spergel}, whereas this was not the case for the 
Gold sample.  In our case, the Gold sample prefers a closed constant-$q$ 
model, whereas the SNLS sample prefers a flat constant-$q$ model.  
Due to the more homogeneous nature of the SNLS sample, the authors of 
\cite{jassal} prefer this sample.  For both cases, however, our study 
shows that there is little model-independent information to be 
extracted from the data beyond the fact that the universe is 
accelerating.

\subsection{Redshift of transition from deceleration to acceleration}

\begin{figure}[ht!]
\includegraphics[width=8cm,height=8cm]{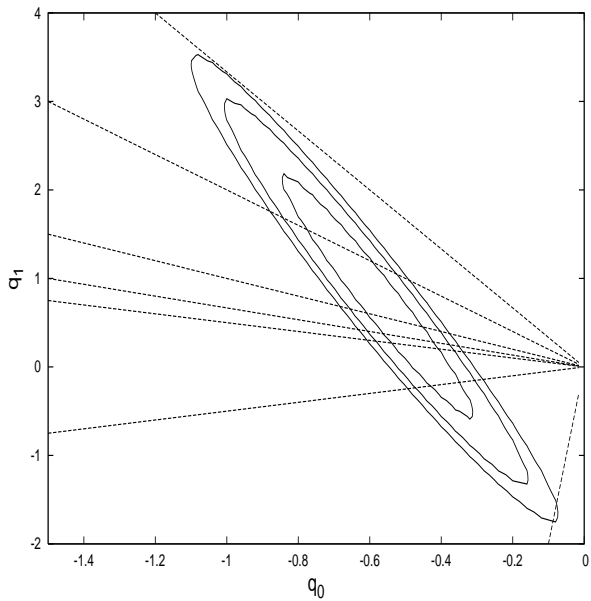}
\caption{Likelihood contours in the $q_0$-$q_1$ plane for the fit of 
model $M_1$ to the SNLS data. The straight lines are lines of 
constant $z_{\rm t}$, with $z_{\rm t}$ varying between $-2$ and $2$.}
\label{fig:q0q1snls}
\end{figure}

Riess et al. used the `Gold' sample to constrain the redshift of 
transition from deceleration to acceleration, finding a transition at 
$z_{\rm t}=0.46\pm 13$ \cite{riess}.  This result was obtained using a linear 
expansion of $q(z)$, the model we have labelled $M_1$.  As pointed out 
by Shapiro and Turner \cite{shapiro}, this model {\it always} has 
a transition redshift at $z_{\rm t} = - q_0 / q_1$, provided $q_0$ and 
$q_1$ have opposite signs, so one should not assign too much significance 
to it.  In particular, the best-fit values of $q_0$ and $q_1$ for a 
given data set might easily give a value of $z_{\rm t}$ outside of 
the redshift range of the SNIa sample used, and $z_{\rm t}$ is then 
clearly unphysical.  We find this to be the situation with the 
SNLS sample.  The best-fit values of $q_0$ and $q_1$ in table 
\ref{tab:resultsSNLS} give $z_{\rm t}\approx 2.0$, while the 
SNLS sample only extends out to $z\sim 1$.  That the linear expansion 
is not a good way of constraining $z_{\rm t}$ is also evident from the 
likelihood contours in the $q_0$-$q_1$ plane, shown in figure 
\ref{fig:q0q1snls}.  The data are consistent with a wide range of 
values of $z_{\rm t}$, even negative values are allowed.  
A better way of looking for a transition from deceleration to acceleration 
is by way of our model $M_5$: parametrizing $q(z)$ as a piecewise constant 
in two bins with the redshift of transition as an additional parameter 
in the fit.  However, since neither the SNLS nor the `Gold' sample 
favour this model, it is at the moment not possible to say anything 
about when (or, indeed, if) the Universe went from deceleration to 
acceleration.  

\subsection{Prior selection}
In general, prior selection is a crucial aspect of fitting model parameters 
and maybe even more when calculating the Bayesian Evidence. 
From the definition, Eq. (\ref{be}), we see that the larger
the volume of the prior parameter space, the smaller the Bayesian
Evidence will be due to normalization. In other words Bayesian Evidence
penalizes a model for large parameter priors, which sounds reasonable
since then model with a large uncertainty in parameter space will
typically be classified as less significant as the same model but
with tighter priors.

\begin{figure}[ht]
\includegraphics[width=50mm,angle=-90]{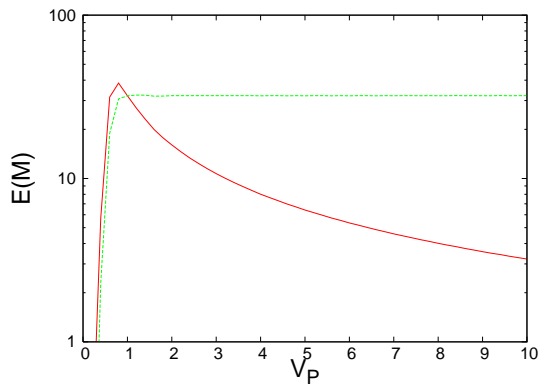}
\caption{Bayesian evidence $E(M)$ (solid red line) along with the
unnormalized Bayesian evidence (green dotted line) the constant $q$ model
as a function of the prior volume $V_p$. Normalization is arbitrary.}
\label{fconstq}
\end{figure}

As an example, consider the Bayesian Evidence for the constant $q$ model
depicted in Fig. \ref{fconstq}(a). The (red) solid line is the BE
as defined in Eq. (\ref{be}) and the (green) dashed line is the unnormalized
Bayesian Evidence, Eq. (\ref{ube}). 
The prior volume in this case is just the difference
between the maximum and minimum $q_0$ value centered around $q_0=0$,
ie. the horizontal axis represents the size of the prior.
For this model, the maximum likelihood point, or minimum $\chi^2=182.8$ 
was found at $q_0=-0.29$. From the figure we see
that the Bayesian Evidence first grows quickly with increasing prior
volume, $V_p$, then reaches a maximum and after that begins to decrease. The
unnormalized BE behaves similarly until the maximum value at which saturates
to a constant. This behavior is simple to understand: when the
prior is too small so that is does not encompass the point
of maximum likelihood, the BE is relatively small. The evidence then grows
with the prior volume since the likelihood increases as more
likely points are included within the prior. As the prior becomes large enough
to encapsulate the volume where the likelihood is concentrated, the 
normalization factor becomes more important and the evidence begins to 
decrease. 
As the unnormalized likelihood does not penalize for the size of the 
prior volume, it simply grows until all of the significant likelihood
is within the integral and then asymptotes to a constant.

The same quantitative behaviour is also found when considering more 
complicated models. It is hence clear that the prior volume
and whether the maximum likelihood point is included within it, are
critical when determining the Bayesian Evidence of a model. If the 
prior is chosen in such a way the maximum likelihood point is not included,
the evidence for that model may be orders of magnitude smaller than 
that of the same model with somewhat larger priors.
This may lead to incorrect conclusions in comparing models, if
one views the parameter priors as separate from the actual functional
form of the model. 

\begin{figure}[ht!]
\includegraphics[width=50mm,angle=-90]{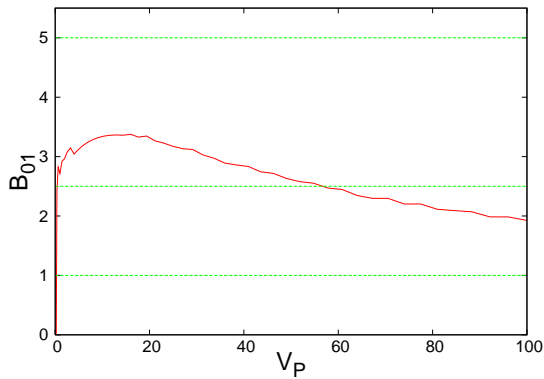}
\caption{
Bayes factor $B_{01}$ between a constant and linear $q$
models for different prior volumes of the linear model. 
The horizontal lines mark the significance levels according to the 
Jeffrey's scale.}
\label{commplot}
\end{figure}

In order to demonstrate how the conclusions may depend on
the prior, we have plotted $\ln B_{01}$ for different prior volumes
for the linear model in Fig. \ref{commplot}.
From the plot we see that conclusions can slightly change
with different priors but radically different results can only
arise if the prior of the linear model can be restricted to a very
small volume.

\section{Conclusions}

We have subjected the state-of-the art SNIa data to a Bayesian analysis, 
assuming an isotropic and homogeneous universe, but 
making no assumptions about the theory of gravity or the matter-energy 
content of the universe.  The main results are that the two samples do not 
enable us to draw strong conclusions about the underlying model, but that 
there is no evidence that anything beyond a constant, negative 
deceleration parameter is required in order to describe the data.  
This is consistent with the conclusions drawn in \cite{shapiro} based 
on a Principle Components Analysis of the Gold sample.  We think these 
results are important to bear in mind when assessing the relevance of 
fashions like `phantom energy'.  

\acknowledgments

The work of {\O}E is supported by the Research Council of Norway, 
project numbers 159637 and 162830. TM is supported by the Academy of
Finland.



\end{document}